\documentclass[]{JHEP3}

\usepackage{bbm,amsthm,amsmath,cancel}
\usepackage{graphics}

\DeclareMathOperator{\ft}{FT}
\newcommand{\mbf}[1][k]{\mathbf{#1}}

\makeatletter
\let\simpref\old@ref
\makeatother

\title{A new method for calculating the primordial bispectrum in the squeezed limit}
\author{Jonathan Ganc\\ Texas Cosmology Center and Department of Physics, The University of Texas at
Austin, Austin, TX 78712, USA \\ E-mail: \email{jonganc@physics.utexas.edu}}
\author{Eiichiro Komatsu\\ Texas Cosmology Center and Department of Astronomy, The University of Texas at
Austin, Austin, TX 78712, USA \\ E-mail: \email{komatsu@astro.as.utexas.edu}}
\abstract{In 2004, Creminelli and Zaldarriaga proposed a consistency relation for the primordial curvature perturbation of all single-field inflation models; it related the bispectrum in the squeezed limit to the spectral tilt. We have developed a technique, based in part on the Creminelli and Zaldarriaga argument, that can greatly simplify the calculation of the squeezed-limit bispectrum using the in-in formalism; we were able to arrive at a generic formula that does not rely on a slow-roll approximation. Using our formula, we explicitly tested the consistency relation for power-law inflation and for an exactly scale-invariant model by Starobinsky; for the latter model, Creminelli and Zaldarriaga's argument predicts a vanishing bispectrum whereas our quantum calculation shows a non-zero bispectrum that approaches zero in the long-wavelength limit and for inflation with a large number of e-folds.}
\preprint{[arXiv:1006.5457]}

\begin{document}
\section{Introduction}
\label{sec:intro}

Inflation was originally proposed in the 1980's to solve the monopole problem, the horizon problem, and the flatness problem \cite{Starobinsky:1979ty,Sato:1980yn,Guth:1980zm,Linde:1981mu,Albrecht:1982wi}. However, its strongest validation has come from the study of cosmic perturbations, since inflation provides a natural mechanism for producing the large scale perturbations observed in the CMB and in large-scale structure \cite{Mukhanov:1981xt,Hawking:1982cz,Starobinsky:1982ee,Guth:1982ec,Bardeen:1983qw,Fischler:1985ky}. We now know that the fluctuation produced were nearly scale invariant and were of the order of 1 part in $10^5$ of the energy density.

 The observable consequences of inflation are usually written on terms of correlation functions. One of the most promising avenues for further discrimination among models of inflation is the study of non-Gaussianity -- that is, the connected part of the three-point or higher-point correlation functions -- in the primordial curvature perturbation $\zeta$. $\zeta$ is conveniently defined as the scalar perturbation in comoving gauge\footnote{This corresponds to a gauge where $H_T=0$ (in the notation of \cite{Kodama:1985bj}) and, for single-field inflation, $\delta\varphi=0$.}, where the metric including only scalar perturbations is
\begin{align}
\label{eq:metric}
  ds^2 = -(1+A)dt^2 + 2 aB_i dx^i \, dt + 
    a^2(t) e^{2 \zeta(\mbf[x])} dx^2.
\end{align}
At later stages, $\zeta$ (or nearly equivalently, the Newtonian potential $\Phi$) can be used to predict perturbations in large-scale structure or the cosmic microwave background. For single-field inflation, $\zeta$ is particularly useful because it is conserved outside the horizon \cite{Bardeen:1983qw,mukhanov1992,Malda}.  The lowest-order non-Gaussianity is the three-point function or bispectrum, which we parametrize as follows:
\begin{align}
  \left\langle \zeta_{\mbf_1} \zeta_{\mbf_2} \zeta_{\mbf_3} \right\rangle
  = (2 \pi)^3 \, \delta^{(3)}\!\big(\,{\textstyle\sum} \mbf_i\big) \,
    B_\zeta(\mbf_1,\mbf_2,\mbf_3).
\end{align}
For future reference, we will similarly write the power spectrum as 
\begin{align}
\label{eq:power_spectrum}
  \left\langle \zeta_{\mbf_1} \zeta_{\mbf_2} \right\rangle
  = (2 \pi)^3 \delta^3 \left( \mbf_1 + \mbf_2 \right)
    P_\zeta(k),
\end{align}
and define the spectral tilt $n_S$ as $n_s - 1 \equiv \frac{d \ln[k^3 P(k)]}{d \ln k}$, so that we have $P \propto k^{-4 + n_s}$.

  The first results about primordial non-Gaussianity were in terms of the non-linearity parameter $f_{\text{NL}}$ \cite{Komatsu:2001rj} (sometimes called $f_{\text{NL}}^{\text{loc}})$; $f_{\text{NL}}$ remains the best-constrained non-Gaussianity measurement, with $f_{\text{NL}} = 32 \pm 21$. $f_{\text{NL}}$ parametrizes the part of the bispectrum that has the form
\begin{align*}
  \left\langle \zeta_{\mbf_1} \zeta_{\mbf_2} \zeta_{\mbf_3}\right\rangle
    = (2 \pi)^3 \, \delta^{(3)}\!\big(\,{\textstyle\sum} \mbf_i\big) \, \frac{6}{5} f_{\text{NL}}
     [P_\zeta(k_1) P_\zeta(k_2) + P_\zeta(k_2) P_\zeta(k_3) + P_\zeta(k_3) P_\zeta(k_1)]
     \,.
\end{align*}
Since we measure that $P_\zeta(k) \propto k^{-3}$ \cite{Komatsu:2010fb}, it is easy to see that this bispectrum has a squeezed ``shaped'' --- i.e. that it peeks in the squeezed limit --- and we can calculate its squeezed limit form\footnote{Remember that, in the squeezed limit $k_3\ll k_1$ so that  $k_1 \approx k_2$.}
\begin{align}
  \left\langle \zeta_{\mbf_1} \zeta_{\mbf_2} \zeta_{\mbf_3}\right\rangle_{k_3\ll k_1, k_2}
    = (2 \pi)^3 \, \delta^{(3)}\!\big(\,{\textstyle\sum} \mbf_i\big) \, \frac{12}{5} f_{\text{NL}}
     P_\zeta(k_1) P_\zeta(k_3)\,.
\end{align}
Clearly, an important observational test for any potential model of inflation is that its predicted $f_{\text{NL}}$ be compatible with observed results.

In 2004, Creminelli and Zaldarriaga \cite{creminelli04}, generalizing an observation from an earlier paper by Maldacena \cite{Malda}, used a clever argument to impose a consistency relation on the three-point function of single-field inflation, or more specifically, on the three-point function in the local limit (i.e. when one of the wavenumbers is considerably smaller than the other two). They demonstrated, using a purely classical argument, that for any single-field inflation model:
\begin{align}
\label{eq:CoR}
  \left\langle \zeta_{\mbf_1} \zeta_{\mbf_2} \zeta_{\mbf_3}\right\rangle_{k_3\ll k_1, k_2}
    = (2 \pi)^3 \, \delta^{(3)}\!\big(\,{\textstyle\sum} \mbf_i\big) \, (1 - n_s) P_\zeta(k_1) P_\zeta(k_3).
\end{align}
In other words, for any single-field model, the consistency relation tells us
\begin{align}
  f_{\text{NL}} = \frac{5}{12}(1-n_s)\,.
\end{align}
Since $1-n_S = 0.037$ \cite{Komatsu:2010fb}, every single-field model should produce $f_{\text{NL}} \approx 0.02$; it therefore follows that, according to the consistency relation, any measurement of $f_{\text{NL}}$ of order 1 or higher \textit{rules out single-field inflation}. Thus, the consistency relation can potentially help rule out a very general class of models. For this reason, it is important to thoroughly understand it. 

In particular, it is interesting to consider what can be learned from trying to verify the consistency relation using a quantum field theory approach rather than the classical methods of the Creminelli and Zaldarriaga proof. One can see one fruit of this approach when we test the Starobinsky model in subsection \ref{subsec:starob_model}, where we can quantitatively determine corrections to the consistency relation prediction in the case of a finite inflationary period where all modes leave the horizon.

In this paper, we will demonstrate a  technique for calculating the bispectrum in the local limit using the in-in formalism without first calculating the general bispectrum; we will then verify the consistency relation for three models. In section \ref{sec:CR_arg}, we will review the argument in \cite{creminelli04}, since it shares some assumptions with our technique. In section \ref{sec:our_formalism}, we will outline our approach. In the next two sections, we will develop our approach in more detail. First, in section \ref{sec:drvng_acn}, we derive the action we will use to calculate the bispectrum. Then, in section \ref{sec:calc_bispectrum}, we perform steps outlined earlier and arrive at (\ref{eq:main_eq}), the  general formula for the bispectrum. In section \ref{sec:check_CoR}, we use our formalism to explicitly verify the consistency relation for slow-roll inflation (a known result), power-law inflation and an exactly solvable model \cite{Starobinsky:2005ab} by Starobinsky, the latter two with no slow-roll approximation. Finally, in section \ref{sec:discussion} we will discuss the implications of our results as well as mention other situations where our formalism would be useful. In appendix \ref{sec_ap:in_depth_calcs}, we outline some of the manipulations in section \ref{sec:calc_bispectrum} more explicitly; in appendix \ref{sec_ap:properties_of_Hankel_functions}, we list the properties of Hankel functions used in subsection \ref{subsec:power_law_iftn} on power-law inflation.

\section{Review of the Creminelli and Zaldarriaga argument}
\label{sec:CR_arg}

We are interested in calculating $\left\langle \zeta_{\mbf_1} \zeta_{\mbf_2} \zeta_{\mbf_3} \right\rangle$ in the limit that $k_3\ll k_1, k_2$ (so that $k_1 \approx k_2$). Mathematically,
\begin{align}
\label{eq:expv}
  \left\langle \zeta_{\mbf_1} \zeta_{\mbf_2}
    \zeta_{\mbf_3} \right\rangle = \Big\langle \,\big\langle
  \zeta_{\mbf_1} \zeta_{\mbf_2}
  \big\rangle_{\zeta_{\mbf_3}} \zeta_{\mbf_3} \Big\rangle,
\end{align}
where we define $\big\langle \ldots \big\rangle_{\zeta_{\mbf_3}}$ to be the expectation value of $\ldots$ given that $\zeta_{\mbf_3}$ has a particular value. Our focus (both here and in later sections) will revolve around calculating $\big\langle \zeta_{\mbf_1} \zeta_{\mbf_2} \big\rangle_{{\zeta_{\mbf_3}}}$.

We will evaluate $\big\langle \zeta_{\mbf_1} \zeta_{\mbf_2} \big\rangle_{{\zeta_{\mbf_3}}}$ after the $k_1, k_2$ modes have crossed the horizon so that the $k_3$ mode will have crossed the horizon in the distant past. Thus, $\zeta_{\mbf_3}$ will be part of an essentially classical \cite{kiefer,komatsu_th} background $\zeta^B$ which affects the scalar field through the metric. Considering only modes far outside the horizon, the metric is
\begin{align}\label{eq:mc_OHZ}
  ds^2 = - dt^2 + a^2(t) e^{2 \zeta^B(\mbf[x])} dx^2,
\end{align}
where
\begin{align*}
  {\zeta}^B(\mbf[x]) \equiv \int_{k\ll k_1, k_2} \frac{d^3k}{(2 \pi)^3}
   {\zeta}_{\mbf} 
    e^{i \mbf\cdot \mbf[x] }.
\end{align*} 
(Note that this also fixes our Fourier convention).

For the next step, let us consider the equivalent real space correlation $\big\langle \zeta^2\big\rangle_{\zeta^B}(\mbf[x]_1,\mbf[x]_2)$ (we will use $\big\langle \ldots \big\rangle_{\zeta^B}$ instead of  $\big\langle \ldots \big\rangle_{\zeta_{\mbf_3}}$ for the remainder of this section\footnote{Otherwise, \eqref{eq:power_series} becomes very confusing.}). We know that the background perturbation $\zeta^B$ is small, so it makes sense to expand the correlation function in a power series about $\zeta^B$ and keep only the first term:
\begin{align}
\label{eq:power_series}
  \big\langle \zeta^2\big\rangle_{\zeta^B}(\mbf[x],\Delta\mbf[x])
   = \big\langle \zeta^2\big\rangle_0(\Delta x)
    + \int d^3k \zeta^B_{\mbf} \left.
     \left(\frac{\delta}{\delta{\zeta}_{\mbf} }
     \right|_{{\zeta}^B=0} 
     \big\langle \zeta^2 \big\rangle_{{\zeta}^B}
       (\mbf[x],\Delta\mbf[x])\right)+\ldots \; ,
\end{align}
where $\mbf[x] \equiv (\mbf[x]_1 + \mbf[x]_2)/2$ and $\Delta\mbf[x] \equiv \mbf[x]_2 - \mbf[x]_1$.

To evaluate this expression, we need to know more about $\big\langle \zeta^2 \big\rangle_{\zeta^B}$. For $\mbf_3$ small enough, we can perform a coordinate change $\mbf[x] \to \mbf[x]' = e^{\zeta^B(\mbf[x])} \mbf[x]$ to put (\ref{eq:mc_OHZ}) in the form of the unperturbed FRW metric. In these new coordinates, the background is unperturbed so
\begin{align}
  \big\langle \zeta^2 \big\rangle_{{\zeta}^B} (\mbf[x],\Delta\mbf[x])
   \approx \big\langle \zeta^2 \big\rangle_0(|\mbf[x]'_2-\mbf[x]'_1|)
   \approx \big\langle \zeta^2 \big\rangle_0(e^{{\zeta}^B(\mbf[x])} \Delta x).
\end{align}
Thus,
\begin{align}
  \left.\frac{\delta}{\delta{\zeta}_{\mbf} }\right|_{{\zeta^B}=0}
    \big\langle \zeta^2 \big\rangle_{{\zeta}^B} (\mbf[x],\Delta\mbf[x])
   = \frac{e^{i \mbf\cdot \mbf[x]}}{(2 \pi)^3}
    \frac{d \left[\big\langle \zeta^2 \big\rangle_0(\Delta x)\right]}{d \ln \Delta x};
\end{align}
substituting this into \eqref{eq:power_series}, moving to Fourier space, and correlating with $\zeta_{\mbf_3}$:
\begin{align*}
 \left\langle \zeta_{\mbf_1} \zeta_{\mbf_2}
   \zeta_{\mbf_3} \right\rangle
  &=\Big\langle \,\big\langle
   \zeta_{\mbf_1} \zeta_{\mbf_2}
   \big\rangle_{\zeta_{\mbf_3}} \zeta_{\mbf_3} \Big\rangle \cr
  &= (2 \pi)^3 \, \delta^{(3)}\!\big(\,{\textstyle\sum} \mbf_i\big) \, P(k_3)
    \ft \left[\frac{d \big\langle \zeta^2 \big\rangle_0}{d \ln \Delta x}\right](k_S) \cr
  &= -(2 \pi)^3 \, \delta^{(3)}\!\big(\,{\textstyle\sum} \mbf_i\big) \, P(k_3)
   \frac{d \left[ k_S^3 P(k_S) \right] }{d \ln k_S} \\
  &= (2 \pi)^3 \, \delta^{(3)}\!\big(\,{\textstyle\sum} \mbf_i\big) \, P(k_1) P(k_3)
   (1-n_S)\,,
\end{align*}
thus yielding the result of \cite{creminelli04}. Note that $\mbf_S\equiv (\mbf_1 - \mbf_2) /2 \approx \mbf_1$.

\section{Outline of our formalism}
\label{sec:our_formalism}

In this section, we outline our approach, which we then follow in more detail over the next 2 sections.
As in the proof in \cite{creminelli04}, we will calculate $\big\langle \zeta^2 \big\rangle_{\zeta_{\mbf[k]_3}}$ and then use $\big\langle \zeta^3 \big\rangle= \big\langle \big\langle \zeta^2 \big\rangle_{\zeta_{\mbf[k]_3}} \zeta_{\mbf[k]_3} \big\rangle$. We will also split $\zeta$ into 2 parts: a large-scale, classical, background part $\zeta_L$ and a small-scale part $\zeta_S$ which undergoes quantum fluctuations; we let
\begin{align}
  \zeta_L \equiv \int_{k < k_*} \frac{ d^3 k}{(2 \pi)^3} \zeta_{\mbf}
    e^{i \mbf \cdot \mbf[x]}, 
   \qquad \zeta_S \equiv \int_{k > k_*}
    \frac{d^3 k}{(2 \pi)^3} \zeta_{\mbf} e^{i \mbf \cdot \mbf[x]},
\end{align}
(where we choose $k_*$ such that $k_3 < k_* \ll k_1,k_2$), so that $\zeta = \zeta_L + \zeta_S$. Note that our method is distinct from stochastic inflation \cite{Starobinsky:1986fx} because we use canonical quantization. Also, we do not worry about the precise value of the cutoff $k_*$ (an issue which takes on some significance in Stochastic inflation; see, for example \cite{Habib:1992ci}) because we are performing our calculation near the time that $k_1,k_2$ exit the horizon, so that $\zeta_L$ (which has $k<k_*\ll k_1$) is almost completely classical \cite{kiefer,komatsu_th}.

To calculate $\big\langle \zeta^2 \big\rangle_{\zeta_{\mbf[k]_3}}$ accurately, we will use cosmological perturbation theory. We start with the Lagrangian for a single scalar-field $\phi$ and we work in a flat Friedman-Robertson-Walker (FRW) background. By an appropriate gauge choice, we can take $\zeta$ to be our degree of freedom (instead of $\phi$); $\zeta$ is more convenient for calculations since it is conserved outside the horizon. Since we ultimately want the three-point vertex $\big\langle \zeta^3 \big\rangle$, we need the Lagrangian up to third order in $\zeta$. Then, we split $\zeta$ in the Lagrangian, using $\zeta = \zeta_L + \zeta_S$, and keep terms of order $\zeta_S^2$ and $\zeta_S^2 \zeta_L$. Also, we will only need the terms that are lowest order in spacial derivatives and in time derivatives, since these will dominate in the squeezed limit; this will allow us to drop a number of terms. If the space and time derivatives on a term act in opposite ways it may not be clear how such a term compares to other terms and we have to keep it\footnote{For example, if we compare a term like  $\partial^{-2} \dot \zeta$ (where $\partial^{-2}$ is the inverse Laplacian) with $\zeta$, the first term has 1 time derivative but -2 space derivatives while the second has 0 time derivatives and 0 space derivatives. Thus, we can not say which one dominates and we need to keep both. This problem actually only comes up once and it turns out we can safely ignore that term.}. 

We can now canonically quantize $\zeta_S$. We treat the terms of order $\zeta_S^2$ as providing the equation of motion for the mode functions of $\zeta_S$ and we consider the terms of order $\zeta_S^2 \zeta_L$ as being perturbations. Then, we use the in-in formalism (a technique in quantum perturbation theory useful in cosmology) to calculate $\big\langle \zeta_S^2 \big\rangle_{\zeta_{\mbf[k]_3}}$. Finally, we correlate with $\zeta_{\mbf[k]_3}$ and arrive at $\big\langle \zeta^3 \big\rangle$.

From this point forward, we will sometimes omit the subscript on  $\left\langle \ldots \right\rangle_{\zeta_{\mbf[k]_3}}$, except where it adds clarity.

\section{Deriving the action}
\label{sec:drvng_acn}

In this section, we will essentially follow \cite{Malda}.

We will work in units where $c=\hbar\equiv1$ and $m_{\text {Pl}}^{-2}=8 \pi G_N\equiv1$. We assume that, on average, inflation takes place on a homogeneous, nearly flat background; thus, we have the following FRW background metric:
\begin{align}\label{eq:FRW_mc}
  ds^2 = - dt^2 + a^2(t) dx^2.
\end{align}
It is sometimes convenient to use conformal time $\eta$, where $\eta \equiv \int dt\,a^{-1}$ so that $dt = a\,d\eta$. Dots over characters (e.g $\dot \phi$) refer to derivatives with respect to physical time $t$, while primes (e.g $\phi'$) refer to derivatives with respect to conformal time. We define the usual Hubble parameter as $H\equiv \dot a / a$.

We suppose that a single scalar-field $\phi$ with a canonical kinetic term is the inflaton\footnote{The generalization to a non-standard kinetic term is done in \cite{RenauxPetel:2010ty}}. The action for $\phi$ is
\begin{align*}
  S = \frac{1}{2} \int d^4 \sqrt{-g }x [R - (\nabla \phi)^2 - 2 V(\phi)].
\end{align*}
We write $\phi(\mbf[x]) = \phi_0(t) + \delta\phi(\mbf[x])$, where $\phi_0$ is the homogeneous background part of the field and $\delta\phi$ is a small perturbation. Then, from equations of motion and conservation of energy, the background $\phi_0$ obeys the following:
\begin{subequations}\label{eq:sclr_fld_eqtns}
  \begin{gather}
    3 H^2=\frac{1}{2}\dot{\phi_0}^2+V(\phi_0),\\
    \dot{H}=-\frac{1}{2}\dot{\phi_0}^2,\\
    0=\ddot{\phi_0}+3H\dot{\phi_0}+V'(\phi_0).
  \end{gather}
\end{subequations}
To determine the dynamics of perturbations, we need to calculate the action for $\delta\phi$ (or equivalently, for $\zeta$). The most efficient way of doing this is to use the Arnowitt-Deser-Misner formalism \cite{Arnowitt:1962hi}, where we write the metric as
\begin{align*}
  ds^2 = - N^2 dt^2 + h_{ij} (dx^i + N^i dt) (dx^j + N^j dt).
\end{align*}
The purpose of the ADM formalism is that we can treat the lapse, shift variables $N, N^i$, respectively, as Lagrange multipliers in order to integrate out their degrees of freedom. Using these variables, we can rewrite the action as
\begin{align*}
  S = \frac{1}{2} \int d^4x \sqrt{h} 
    \left[ N R^{(3)} - 2 N V + N^{-1} (E_{ij} E^{ij} - E^2) 
      + N^{-1} ( \dot \phi - N^i \partial_i \phi)^2 - N h^{ij} \partial_i \phi \partial_j \phi \right],
\end{align*}
where
\begin{align*}
  E_{ij} & = \frac{1}{2} ( \dot h_{ij} - \nabla_i N_j - \nabla_j N_i),\cr 
  E & = E_i^i.
\end{align*}
where the covariant derivatives on $N_i$ are taken with respect to the three-geometry. 

We will adopt comoving gauge \cite{Kodama:1985bj,mukhanov1992} where, considering only scalar perturbations, 
\begin{align*}
  \delta\phi = 0, \qquad h_{ij} = a^2 e^{2 \zeta} \delta^{ij}.
\end{align*}
In this gauge, we have shifted the degree of freedom from $\delta\phi$ to $\zeta$. $\zeta$ is the same quantity introduced early, namely the Bardeen variable that is conserved outside the horizon \cite{Wands:2000dp,Lyth:2005du}.

We can now exploit the power of the ADM formalism. We find the equations of motion for $N^i$ and $N$, solve them to the necessary order in $\zeta$ ($2^\text{nd}$ order for our purposes), and then replace them in the action\footnote{Again, this is done more explicitly in \cite{Malda}.}.

Up to third order in $\zeta$, we find
\begin{align}
 \label{eq:full_acn}
 &S\phantom{_2}=S_2 + S_3,\cr
 &S_2 = \frac{1}{2} \int d^4x \frac{\dot \phi_0}{H^2} 
  [a^3 \dot \zeta^2 - a (\partial \zeta)^2],\cr
 &\begin{aligned}
  S_3 =
   & \int d^4x 
    \frac{1}{4} \frac{\dot \phi_0^4}{H^4} [ a^3 \dot \zeta^2 \zeta + a (\partial \zeta)^2 \zeta ] 
     - \frac{ \dot \phi_0^2}{H^2} a^3 \dot \zeta \partial_i \chi \partial_i \zeta \cr
   &  -\frac{1}{16} \frac{ \dot \phi_0^6}{H^6} a^3 \dot \zeta^2 \zeta
    + \frac{\dot \phi_0^2}{H^2 } a^3 \dot \zeta \zeta^2 
     \frac{d}{dt} \left[\frac{1}{2}  \frac{\ddot \phi_0}{\dot \phi_0 H }
       + \frac{1}{4} \frac{\dot \phi_0^2}{H^2} \right]
    + \frac{1}{4} \frac{ \dot \phi_0^2}{H^2 } a^3 \partial_i \partial_j \chi \partial_i \partial_j \chi \zeta \cr
   & - \left[ \frac{1}{2} \frac{ \ddot  \phi_0}{\dot \phi_0 H} \zeta^2 
    + \frac{1}{4} \frac{ \dot \phi_0^2}{H^2 } \zeta^2 +\right.  \cr 
   &~~~~ + \frac{ 1}{H} \dot \zeta \zeta
    -\frac{1}{4} \frac{ a^{-2}}{H^2 } (\partial \zeta)^2
    + \frac{ 1}{ 4} \frac{ a^{-2}} {H^2 } \partial^{-2} \partial_i \partial_j ( \partial_i \zeta \partial_j \zeta)
    + \frac{1}{2} \frac{ 1}{H} \partial_i \chi \partial_i \zeta \cr
   &\left.~~~~ 
     - \frac{1}{2} \frac{ 1}{ H} \partial^{-2} \partial_i \partial_j ( \partial_i  \chi \partial_j \zeta) \right]
    \left.\frac{ \delta L}{\delta \zeta}\right|_1,
 \end{aligned}\cr\vspace{1mm}
\end{align}
where 
\begin{align*}
\left.\frac{ \delta L}{\delta \zeta}\right|_1 = \frac{ \delta S_2}{\delta \zeta}= 
  - \partial_t ( a^{ 3} \frac{ \dot \phi_0^2}{H^2} \dot \zeta ) 
    +  a \frac{ \dot \phi_0^2}{H^2} \partial^2 \zeta\,;
\end{align*}
to lowest order, the equation of motion for $\zeta$ is $\left.\frac{ \delta L}{\delta \zeta}\right|_1=0$. 

This is the action we want. Note that we have not made any slow-roll assumptions.

\section{Calculating the local-limit bispectrum}
\label{sec:calc_bispectrum}

As described in section \ref{sec:our_formalism}, our goal is essentially to determine $\big\langle \zeta_{\mbf[k]_1} \zeta_{\mbf[k]_2} \big\rangle_{\zeta_{\mbf[k]_3}} = \big\langle \zeta_{S,\mbf[k]_1} \zeta_{S,\mbf[k]_2} \big\rangle_{\zeta_{\mbf[k]_3}}$, that is, the two-point function calculated with a classical background $\zeta^B$, where $\zeta_{\mbf[k]_3}^B = \zeta_{\mbf[k]_3}$. Ordinarily --- i.e. if we do not fix a perturbed background ---  $\big\langle  \zeta_{\mbf[k]_1} \zeta_{\mbf[k]_2} \big\rangle \propto \delta(\mbf_1 + \mbf_2) = 0$ if $\mbf_2\neq -\mbf_1$. Thus, the two-point function $\big\langle \zeta^2 \big\rangle$ is only non-zero insofar as it is affected by  a contribution from the background, which arises from  the third-order term in (\ref{eq:full_acn}). We will see that the contribution has two pieces: one from the field redefinition of $\zeta_S$ in (\ref{eq:fld_rdfnn}) and one from the in-in formalism.

\subsection{Finding the action in terms of $\zeta_S$ and $\zeta_L$}
\label{subsec:acn_zs_zl}

Using $\zeta=\zeta_S + \zeta_L$, we find the terms of order $\zeta_S^2$ in $S_2$, which we call $S_0$, and the terms of order $\zeta_S^2 \zeta_L$ in $S_3$, which we call $S_\text{int}$\footnote{There are terms of order $\zeta_S \zeta_L$ which could, in principle, be  significant. However, after an integration by parts, they become equal to
\begin{align*}
  \int d^4x \, \zeta_S \left.\frac{ \delta L}{\delta \zeta}\right|_{\zeta =
    \zeta_L}.
\end{align*}
Since $\zeta_L$ is a classical field it obeys its equations of motion, so $\left.\frac{ \delta L}{\delta \zeta}\right|_{\zeta = \zeta_L}=0$ and these terms do not contribute.

Any terms of order $\zeta_S \zeta_L^2$ can also be ignored  because they can not contribute to a two-point correlation $\left\langle \zeta_S^2 \right\rangle$}%
. $S_0$ will determine the equation of motion for $\zeta_S$, and $S_\text{int} $ will be the perturbation to $S_0$ and will enter into the in-in formalism to calculate $\big\langle \zeta^2 \big\rangle$.

From $S_2$, we see that
\begin{align}\label{eq:s0}
  S_0= \frac{1}{2} \int d^4x \frac{\dot \phi_0^2}{H^2} 
  [a^3 \dot \zeta_S^2 - a (\partial \zeta_S)^2].
\end{align}
We can similarly pick out the $\zeta_S^2 \zeta_L$ terms in $S_3$. Note first that the two terms with coefficients of $a^{-2}$ have derivatives on all the $\zeta$'s and thus must yield a term with a derivative on $\zeta_L$; these two terms will thus be negligible. From the remaining terms, we substitute in for $\zeta$ and pick out terms with the fewest derivatives on $\zeta_L$\footnote{As mentioned in section \ref{sec:our_formalism}, we retain terms where the effect of the derivatives is unclear, as in the $\partial_i \partial^{-2} \dot \zeta_L$ term. Actually, it turns out that this term does not contribute, though this is not immediately obvious. The reasons are mentioned later in this subsection and in appendix \ref{sec_ap:in_depth_calcs}.}, yielding
\begin{align}
\label{eq:sint-init}
 S_{\text{int}} =  & \int d^4 x\Bigg[\left(\frac{ 1}{ 4 } \frac{\dot \phi_0^4}{H^4} 
     -\frac{ 1}{16} \frac{ \dot \phi_0^6}{H^6} \right) a^3 \zeta_L \dot \zeta_S^2
   + \frac{ 1}{ 4 } \frac{ \dot \phi_0^4}{H^4} a \zeta_L ( \partial\zeta_S)^2 
   - \frac{ \dot \phi_0^4}{2 H^4} a^3 \dot \zeta_S \partial_i \zeta_S \partial_i \partial^{-2} \dot \zeta_L +\cr
 &~~~ + \frac{ 1}{16} \frac{ \dot \phi_0^6}{H^6 } 
    a^3 \partial_i \partial_j \partial^{-2} \dot \zeta_S \, \partial_i \partial_j \partial^{-2} \dot \zeta_S \, \zeta_L
   + 2 \frac{\dot \phi_0^2}{H^2 } a^3 \zeta_L 
    \frac{d}{dt} \left[\frac{ 1}{2}  \frac{\ddot \phi_0}{\dot \phi_0 H } 
     + \frac{ 1}{4} \frac{\dot \phi_0^2}{H^2} \right] \dot \zeta_S \zeta_S \cr
 &~~~ - f(\zeta) \frac{ \delta L_0}{\delta \zeta_S}\Bigg]\,,
\end{align}
 where
\begin{align*}
 f(\zeta) \equiv
  \frac{ \ddot  \phi_0}{\dot \phi_0 H}\zeta_L \zeta_S
  + \frac{ 1}{2} \frac{ \dot \phi_0^2}{H^2 } \zeta_L \zeta_S
  + \frac{ 1}{H} \zeta_L \dot \zeta_S
  + \frac{\dot \phi_0^2}{4 H^3} \partial_i \partial^{-2} \dot \zeta_L \partial_i \zeta_S -
   \frac{\dot \phi_0^2}{4 H^3} 
    \partial^{-2} \partial_i \partial_j ( \partial_i \partial^{-2} \dot \zeta_L \partial_j \zeta_S).
\end{align*}

The terms with a factor of ${ \delta L_0}/{\delta \zeta_S}$ are best removed by a field redefinition:
\begin{align} \label{eq:fld_rdfnn}
  \zeta_S &\equiv \zeta_N + f(\zeta_N)\cr
   &= \zeta_N 
    + \left(\frac{ \ddot  \phi_0}{\dot \phi_0 H}
    + \frac{ 1}{2} \frac{ \dot \phi_0^2}{H^2 } \right)\zeta_L \zeta_N + \ldots,
\end{align}
where, in the second line, 1) we have dropped terms that vanish outside the horizon, since the terms in the field redefinition will only be evaluated outside the horizon (as we will see in subsection \ref{subsec:fld_rdfnn})\footnote{$\dot \zeta$ always goes to zero at late times since $\dot \zeta\propto H^2/a^3\dot \phi_0^2$ far after horizon crossing \cite{mukhanov1992}.}; and 2) we have replaced $\zeta_S$ with $\zeta_N$ since they are effectively equal when they multiply terms of order $\zeta^2$. 

Thus, after the redefinition, our action $S_\text{int}$ in terms of $\zeta_N$ becomes
\begin{align}
\label{eq:sint}
 S_{\text{int}} =  & \int d^4x\Bigg[ \left(\frac{ 1}{ 4 } \frac{\dot \phi_0^4}{H^4} 
     -\frac{ 1}{16} \frac{ \dot \phi_0^6}{H^6} \right) a^3 \zeta_L \dot \zeta_N^2
   + \frac{ 1}{ 4 } \frac{ \dot \phi_0^4}{H^4} a \zeta_L ( \partial\zeta_N)^2 
   - \frac{ \dot \phi_0^4}{2 H^4} a^3 \dot \zeta_N \partial_i \zeta_N \partial_i \partial^{-2} \dot \zeta_L +\cr
 &~~~ + \frac{ 1}{16} \frac{ \dot \phi_0^6}{H^6 } 
    a^3 \partial_i \partial_j \partial^{-2} \dot \zeta_N \, \partial_i \partial_j \partial^{-2} \dot \zeta_N \, \zeta_L
   + 2 \frac{\dot \phi_0^2}{H^2 } a^3 \zeta_L 
    \frac{d}{dt} \! \left(\frac{ 1}{2}  \frac{\ddot \phi_0}{\dot \phi_0 H } 
     + \frac{ 1}{4} \frac{\dot \phi_0^2}{H^2} \right) \dot \zeta_N \zeta_N \Bigg]\,.\cr
\end{align}

\subsection{Field redefinition}
\label{subsec:fld_rdfnn}

$\zeta_S$, rather than $\zeta_N$, is what will remain constant outside the horizon and what we correlate with $\zeta_L$, so we need to calculate $\big\langle \zeta_{S,\mbf[k]_1} \zeta_{S,\mbf[k]_2} \big\rangle$. To understand the effect of a field redefinition, consider a simple case, where $\zeta_S = \zeta_N + \kappa$ and $\kappa\ll \zeta_N$: then 
\begin{align*}
  \big\langle \zeta_{S,\mbf[k]_1}\!(t)\, \zeta_{S,\mbf[k]_2}\!(t)\big\rangle 
   \approx \big\langle \zeta_{N,\mbf[k]_1}\!(t)\, \zeta_{N,\mbf[k]_2}\!(t) \big\rangle 
    + 2 \big\langle \kappa (t)\, \zeta_N\!(t)\big\rangle.
\end{align*}
We actually have (\ref{eq:fld_rdfnn}), where $\zeta_S(x) \approx \zeta_N(x) + \beta(t) \zeta_L(x) \zeta_N(x)$; in this case, we find\footnote{$\ft[f(\mbf[x])](\mbf)$ is defined as
  \begin{align*}
    \ft[f(\mbf[x])](\mbf)
     \equiv \int d^3x f(\mbf[x]) e^{-i \mbf\cdot\mbf[x]}\,.
  \end{align*}}\begin{align*}
  \big\langle \zeta_{S,\mbf[k]_1}\!(t)\, \zeta_{S,\mbf[k]_2}\!(t)\big\rangle 
   \approx \big\langle \zeta_{N,\mbf[k]_1}\!(t)\, \zeta_{N,\mbf[k]_2}\!(t)\big\rangle 
    + 2 \beta(t)\,\zeta_{L,\mbf_1+\mbf_2}\!(t) 
     \,\ft\left[ \big\langle \zeta^2 \big\rangle_0\!(\Delta x,t) \right]\!(k_1)
\end{align*}
(this is worked out in appendix \ref{subsec_ap:fld_rdfnn_calc}). Note that $\ft\left[ \big\langle \zeta^2 \big\rangle_0\!(\Delta x,t) \right]\!(k_1)=\left| u_{k_1}(t)\right|^2$, where $u_{k_1}$ is the mode function defined in the following section. If we measure at a time $\bar{t} \gg t_*$, where $t_*$ is the time that $k_1,k_2$ cross the horizon, then $\ft\left[ \big\langle \zeta^2 \big\rangle_0\!(\Delta x,\bar{t}) \right]\!(k)=\left| u_{k}(\bar{t})\right|^2=P(k)$. Thus,
\begin{align}\label{eq:result_of_fld_rdfn}
  \big\langle  \zeta_{S,\mbf[k]_1}\!(\bar{t})\, \zeta_{S,\mbf[k]_2}\!(\bar{t}) \big\rangle 
   \approx \big\langle \zeta_{N,\mbf[k]_1}\!(\bar{t})\, 
     \zeta_{N,\mbf[k]_2}\!(\bar{t})\big\rangle 
    + 2 \left(\frac{ \ddot  \phi_0}{\dot \phi_0 H}
    + \frac{ 1}{2} \frac{ \dot \phi_0^2}{H^2 } \right)\!(\bar{t})
  \,\zeta_{L,\mbf_1+\mbf_2}\,P(k_1)\,.
\end{align}

We still need to determine $\big\langle \zeta_{N,\mbf[k]_1}\!(\bar{t})\, \zeta_{N,\mbf[k]_2}\!(\bar{t})\big\rangle $, which requires that we use quantum perturbation theory. 

Interestingly, the contribution from the field-redefinition is the same as the non-Gaussianity predicted by the ``$\delta N$'' formalism \cite{Starobinsky:1982ee,Starobinsky:1986fxa,Salopek:1990jq,Sasaki:1995aw}; this connection was originally established by Seery and Lidsey \cite{SL05mlt}, who noted that that the field redefinition used in the bispectrum calculation is similar to the transformation from comoving to flat gauge, and Arroja and Koyama \cite{Arroja:2008ga,Koyama:2010xj}, who observed that the gauge transformation on large scales is given by the ``$\delta N$'' formalism. To verify this result, note that in the ``$\delta N$'' formalism, one supposes that $\phi$ is completely Gaussian and that any non-Gaussianity is generated by non-linearities in the relationship between $\zeta$ and $\phi$:
\begin{align}
  \zeta = - \frac{1}{2} \int_{\phi_0}^{\phi = \phi_0 + \delta\phi} d \phi' 
    \left( \frac{\partial \ln H}{\partial \phi'} \right)^{-1};
\end{align}
to second order in $\delta\phi$, this gives $\zeta = \zeta_L - \left( \frac{\partial^2 \ln H}{\partial \phi^2} \right) \zeta_L^2$, where $\zeta_L = - 1/2 \left( {\partial \ln H}/{\partial \phi'} \right)^{-1} \delta\phi$ is linear in $\delta\phi$ and Gaussian.  Then,
\begin{align}
  \left\langle \zeta^3 \right\rangle
    = - (2 \pi)^3 \, \delta^{(3)}\!\big(\,{\textstyle\sum} \mbf_i\big) 
     \, 2  \, \frac{\partial^2 \ln H}{\partial \phi^2} 
     \big[ P_\zeta(k_1) P_\zeta(k_2) + P_\zeta(k_2) P_\zeta(k_3) 
      + P_\zeta(k_3) P_\zeta(k_1) \big],
\end{align}
which gives us, for single-field inflation,
\begin{align}
  \big\langle  \zeta^3 \big\rangle _{k_3\ll k_1, k_2}
    = - (2 \pi)^3 \, \delta^{(3)}\!\big(\,{\textstyle\sum} \mbf_i\big) 
      \, P_\zeta(k_1) P_\zeta(k_3) \;
      2 \left(\frac{\ddot{\phi_0}}{H\dot{\phi_0}} + 
        \frac{1}{2}\frac{\dot{\phi_0}^2}{H^2}\right);
\end{align}
this is the same as the bispectrum contribution from the second term in (\ref{eq:result_of_fld_rdfn}) after we correlate with $\zeta_{\mbf[k]_3}$.

\subsection{Quantizing $\zeta_N$}
\label{subsec:-qnt_zs}

Here, we quantize $\zeta_N$, following the methods of \cite{birell&davies}. We start by expanding $\zeta_N$:
\begin{align}\label{eq:expd_zet}
  \zeta_N(\mbf[x]) = \int \frac{d^3 k}{(2\pi)^3} \zeta_{N,\mbf} e^{i \mbf \cdot \mbf[x]}\,.
\end{align}
From (\ref{eq:s0}), we find the equation of motion for $\zeta_{N,\mbf[k]}$:
\begin{align}\label{eq:EOM}
  -\frac{d\left(a^3 \frac{\dot \phi_0^2}{H^2} \dot \zeta_{N,k}
    \right)}{dt}
   - \frac{\dot \phi_0^2}{H^2} a k^2 \zeta_{N,k} = 0\,.
\end{align}

To quantize $\zeta_N$, we promote $\zeta_{N,\mbf[k]}$ to an operator
\begin{align}\label{eq:qntz_zet_N}
  \zeta_{N,\mbf[k]}(t) = u_k(t) a_{\mbf} + u^*_k(t) a^\dagger_{-\mbf}\,.
\end{align}
$a_{\mbf}, a^\dagger_{\mbf}$ are annihilation, creation operators for $\zeta_N$, respectively, which obey the canonical commutation relation $\big[ a_{\mbf}, a^\dagger_{\mbf'} \big] = (2 \pi)^3 \delta^{(3)}(\mbf - \mbf')$. $u_k,u^*_k$ are two independent solutions to the equation of motion (\ref{eq:EOM}) for $\zeta_N$, normalized to reflect the appropriate vacuum condition for large $k$; $u_k,u^*_k$ are known as the ``mode functions'' of $\zeta_N$. Note that since $\zeta_N$ is the same as $\zeta_S$ to lowest order in $\zeta$, the mode functions for $\zeta_N$ are the same as those for $\zeta_S$ and for $\zeta$.

At this point, we can find the power spectrum of $\zeta_N$ (which is the same as the power spectrum of $\zeta$):
\begin{align*}
  \big\langle \zeta_{\mbf[k]} \zeta_{\mbf[k]'} \big\rangle
   = (2 \pi)^3 |u_k|^2 \delta^{(3)}(\mbf + \mbf'),
  \intertext{so}
  P(k) = |u_k|^2.
\end{align*}

\subsection{Applying the in-in formalism}
\label{subsec:in_in_formalism}

The in-in formalism \cite{Schwinger:1960qe,weinberg_qccc}\footnote{\cite{weinberg_qccc} has a list of other references on the in-in formalism.}, sometimes called the Keldysh-Schwinger formalism, has become an important tool in the calculation of cosmic correlation functions \cite{Malda,SL05sng,SL05mlt,Chen06,SL07,SL09,Arroja:2008ga,Chen09,Arroja:2009pd}. It specifies the expectation value of quantum operators at some time given a known input state in the past. Here, we are interested in calculating $\big\langle \zeta_N(t) \zeta_N(t) \big\rangle$ given that the universe was in a vacuum state in the far past. 

The question of selecting the proper initial vacuum state is non-trivial and has been written about extensively \cite{birell&davies,Greene:2005wk,Chen06}\footnote{\cite{Chen06} has a list of other references on the effect and intricacies of initial vacuum selection.}. 
The effect of vacuum state choice is usually accounted for by mixing the mode functions $u, u^*$ to form new mode functions $\tilde{u}, \tilde{u}^*$. In this section, we do not explicitly specify mode functions, so we are not forced to choose an initial state. In the following section, where we work examples, we choose the standard Bunch-Davies vacuum, which assumes that the mode functions of fields limit toward their standard Minkowski-space form at early times and/or deep inside the horizon.

The formula for the tree level in-in formalism (sufficient for our calculation) is
\begin{align}\label{eq:in-in}
  \big\langle  \zeta^2(\bar{t}) \big\rangle
   = - i \int_{- (1 - i \epsilon)\, \infty}^{\bar{t}} dt' 
    \big\langle 0\big| [\zeta^2(\bar{t}), H_I(t')]\big| 0\big\rangle,
\end{align}
where $H_I$ is the interaction Hamiltonian in the interaction picture and $\big\langle 0\big| \ldots \big| 0\big\rangle$ means that we take the expectation value in the free-field theory. In principle, there is a distinction between variables in the interaction picture and in the Heisenberg picture but this can largely be ignored without any complications (see \cite{wein_qtf1} for more about the interaction picture). A different, but relevant, subtlety is that the time integral contour in (\ref{eq:in-in}) picks up a small imaginary part at early times, so that we have written the lower limit of the integral as $- (1 - i \epsilon)\, \infty$. This contour should be familiar to those who have done standard perturbation calculations in QFT; in both cases, the purpose of the imaginary part is to perform the calculation in the perturbed (rather than the free) vacuum. If the mode functions are in the Bunch-Davies vacuum, this usually makes the integrand exponentially decay at early times so that the integral converges.   

It is usually true, including in this case, that $H_I = - L_\text{int}$. Then, inserting our expression (\ref{eq:sint}) for $L_\text{int}$ into the formula above (see appendix \ref{sec_ap:in_depth_calcs} for the explicit calculation), we arrive at:
\begin{gather}
 \nonumber
 \big\langle \zeta_{N,\mbf[k]_1} \zeta_{N,\mbf[k]_2} \big\rangle_{\zeta_{\mbf[k]_3} }
   = \zeta_{L,\mbf[k]_1 + \mbf[k]_2} K,
  \intertext{where}
 \begin{aligned}\label{eq:dfn_K}
  K \equiv i u_{k_1}^2(\bar{\eta}) 
   \int_{-\infty (1-i\epsilon)}^{\bar{\eta}}  d\eta
    \Bigg[ 
     \frac{ 1}{ 2 }& \frac{ \dot \phi_0^4 }{ H^4} a^2 u_{k_1}'^{*2}(\eta)
      + \frac{ 1}{ 2 } \frac{ \dot \phi_0^4 }{ H^4} a^2 k_1^2
       {u_{k_1}^{*2}}(\eta) + \cr
     &~~+ 2 \frac{\dot \phi_0^2 }{ H^2 } a^3 
       \frac{d }{ dt} \left( \frac{\ddot \phi_0 }{ \dot \phi_0 H }
        + \frac{ 1 }{ 2} \frac{\dot \phi_0^2 }{ H^2} \right)
       {u_{k_1}'^*}(\eta) u_{k_1}^*(\eta)
     \Bigg] + \text{c.c.}\;\;.
 \end{aligned}
\end{gather}

Using this previous formula in (\ref{eq:result_of_fld_rdfn}) and correlating with $\zeta_{\mbf[k]_3}$, we get:
\begin{align}
  \label{eq:main_eq}
  \big\langle \zeta^3 \big\rangle = (2 \pi)^3 \, \delta^{(3)}\!\big(\,{\textstyle\sum} \mbf_i\big) \,
    P(k_3) \left\{ P(k_1) \;
    2 \left( \frac{\ddot \phi_0 }{ \dot \phi_0 H } 
       + \frac{ 1 }{ 2} \frac{\dot \phi_0^2 }{ H^2} \right) + K \right\}.
\end{align}
This is the main result of the paper: a formula for the bispectrum in the squeezed limit, given in terms of the mode functions $u,u^*$. Note again that no slow-roll approximation has been used.

\section{Checking the consistency relation in specific cases}
\label{sec:check_CoR}

\subsection{Slow-roll inflation}
\label{subsec:SwR_iftn}

Slow-roll inflation is the simplest model of inflation and, additionally, is entirely compatible with current observational data. It corresponds to a nearly de Sitter (dS) expansion produced by a single inflaton field. We assume $\dot H = - (1/2) \dot \phi_0^2$ is small and so is $\ddot \phi_0$. More precisely, we define dimensionless \emph{slow-roll parameters}
\begin{subequations} \begin{align}
 \epsilon & \equiv \frac{1}{2}\left(\frac{V'}{V}\right)^2
  \approx -\frac{\dot{H}}{H^2} =\frac{1}{2}\frac{\dot{\phi_0^2}}{H^2},\\
 \tilde{\eta} & \equiv \frac{V''}{V}
  \approx-\frac{\ddot{\phi_0}}{H\dot{\phi_0}} + \frac{1}{2}\frac{\dot{\phi_0}^2}{H^2},
\end{align} \end{subequations}
to be small. We then perform all our calculations to lowest order in these parameters.

In dS space, we can always choose $\eta_0$ and $a_0$ such that $a = - (\eta H)^{-1}$. This is still true for slow-roll inflation, at least to lowest order in slow-roll.

The equation of motion (\ref{eq:EOM}) for $\zeta_N$ and thus for $u$ becomes:
\begin{gather}
  \nonumber
  -\frac{d\left(a^3 \, \dot u_k \right)}{dt} - a k^2 u_k = 0,\cr
  \intertext{so that}
  u_k(\eta)=  \frac{H^2}{\dot \phi_0} \frac{1}{\sqrt{2 k^3}}
    (1+ik \eta) e^{-ik \eta},\\
  \intertext{and}
  \label{eq:SwR-power_spr}
  P_\zeta(k)=|u|^2_{\eta\to0}
    = \frac{H^4}{2 \dot \phi_0^2}\frac{1}{k^3}.
\end{gather}

\paragraph{Spectral tilt}
It is reasonable to suppose that in $P(k)$, ${H^4}/{\dot \phi_0^2} = {H^4}(t_*)/{\dot \phi_0^2}(t_*)$, where $t_*$ is the time that the mode $k$ crosses the horizon, i.e. $k \approx a(t_*) H$. Then, $dk/dt_* \approx a(t_*)H^2 \approx kH$, so $d(\ln k) = H \,dt_*$. Then,
\begin{align}
  n_S - 1 = \frac{d \ln [P(k) k^3]}{\ln k} 
   = \frac{1}{H} \frac{d\ln {H^4}/{\dot \phi_0^2}}{dt_*}
   = -2 (\frac{\ddot{\phi_0}}{H\dot{\phi_0}} 
     + \frac{\dot{\phi_0}^2}{H^2})
   = 2 \eta - 6 \epsilon.
\end{align}

\paragraph{Bispectrum in the squeezed limit}

Using (\ref{eq:main_eq}):
\begin{align}
 \nonumber K &= \frac{i}{2} \frac{\dot \phi_0^4}{H^4} 
  \underbrace{u_{k_1}^2(\bar{\eta})}_{= P(k_1)}
  \int_{-\infty (1-i\epsilon)}^{\bar{\eta}} d\eta
   \Big(a^2 u_{k_1}'^{*2}(\eta) + k_1^2 a^2 u_{k_1}^{*2}(\eta)\Big) 
  + \text{c.c.}\cr
  &\begin{aligned}=  \frac{i}{2} \frac{\dot \phi_0^4}{H^4} P(k_1) 
   \Big[&\frac{H^4}{\dot \phi_0^2} \frac{k_1}{2}
     \int_{-\infty (1-i\epsilon)}^{\bar{\eta}} d\eta
     \Big(\frac{-1}{\eta H}\Big)^2 \eta^2 e^{2 i k_1 \eta}+\cr
    &~+ \frac{H^4}{\dot \phi_0^2} \frac{1}{2 k_1}
     \int_{-\infty (1-i\epsilon)}^{\bar{\eta}} d\eta
     \Big(\frac{-1}{\eta H}\Big)^2 (1-ik_1  \eta)^2 e^{2 i k_1 \eta} \Big]
    + \text{c.c.}\;.
   \end{aligned}
   \intertext{Then, since we take the $\eta\to 0$ limit, to determine the bispectrum at the present we let $e^{2 i k_1 \eta}\to 1+2 i k_1 \eta$:}
  K&=  \frac{i}{2} \frac{\dot \phi_0^4}{H^4} P(k_1) 
   \Big[\frac{H^2}{\dot \phi_0^2} \frac{k_1}{2}
     \Big(\frac{1}{2 i k_1} \Big)
    + \frac{H^2}{\dot \phi_0^2} \frac{1}{2 k_1} k_1
     \Big(\!-\!\frac{1}{k_1 \eta} - \frac{1}{2 i} \Big)
      \Big( 1 - \frac{2 k_1 \eta}{i} \Big)
     \Big]_{\eta=\bar{\eta}}
    + \text{c.c.} \cr
  & = \frac{\dot \phi_0^2}{H^2} P(k_1).
\end{align}
So,
\begin{align}
  \big\langle  \zeta^3 \big\rangle
   &= (2 \pi)^3 \, \delta^{(3)}\!\big(\,{\textstyle\sum} \mbf_i\big) \, 
    P(k_1) P(k_3)\,
     2 \,(\frac{\ddot{\phi_0}}{H\dot{\phi_0}} + \frac{\dot{\phi_0}^2}{H^2})\cr
   &= (2 \pi)^3 \, \delta^{(3)}\!\big(\,{\textstyle\sum} \mbf_i\big) \, 
    P(k_1) P(k_3)\, (1-n_S),
\end{align}
matching the result of \cite{Malda} and verifying the consistency relation.

\subsection{Power-law inflation}
\label{subsec:power_law_iftn}

As another example, consider power-law inflation (where the consistency relation has not previously been verified). It provides an opportunity to test the consistency relation in a fully  non slow-roll situation.

In power-law inflation \cite{Abbott:1984fp,Lucchin:1985wy,Lyth:1992tx,Wein_cosmol}, we suppose $V(\phi_0)=ge^{-\lambda\phi_0}$, where $g$ and $\lambda$ are dimensionless parameters. We can exactly solve the scalar field equations (\ref{eq:sclr_fld_eqtns})  to find
\begin{gather*}
  \phi_0(t) = \frac{1}{\lambda} \ln \left( \frac{g \epsilon^2 t^2}{3-\epsilon} \right)
  \intertext{and}
  H = \frac{1}{\epsilon t};
\end{gather*}
we have defined the dimensionless parameter
\begin{align*}
  \epsilon \equiv - \frac{\dot H}{H^2} = \frac{\lambda^2}{16 \pi G},
\end{align*}
which has some similarities to the slow-roll parameter $\epsilon_{\text{SR}}$ except that here, we only require $\epsilon<1$.

Then, we can choose $a_0$ and $t_0$ such that 
\begin{align*}
  a = t^{1/\epsilon}.
\end{align*}
It is useful to define the quantity
\begin{align*}
  \nu = \frac{3}{2} + \frac{\epsilon}{1-\epsilon};
\end{align*}
then, slow-roll inflation corresponds to $\nu \approx 3/2$.

We can solve (\ref{eq:EOM}) for $u$
\begin{align*}
  u_k(\eta) = \frac{c}{\epsilon} 
   \left( \frac{\epsilon}{1 - \epsilon} \right)^{- \frac{1}{1-\epsilon} }
   \exp \left( \frac{i \pi \nu}{2} + \frac{i \pi}{4} \right)
   (-\eta)^\nu H_\nu^{(1)}(-k \eta),
\end{align*}
where we have defined $c=- \lambda \sqrt{\pi} / 4 (2\pi)^{3/2}$ and $H_\nu^{(1)}(z)$ is a Hankel function; the relevant properties of Hankel functions are listed in appendix \ref{sec_ap:properties_of_Hankel_functions}.

Thus, we find that
\begin{gather}
  P(k) = |u(k)|^2_{-k \eta\ll1} = \frac{c^2}{\epsilon^2}
    \frac{\left[\Gamma(\nu)\right]^2}{\pi^2}
    \left( \frac{\epsilon}{1 - \epsilon} \right)^{- \frac{2}{1-\epsilon} }
    \left( \frac{2}{k} \right)^{2 \nu}
  \intertext{and}
  1-n_S = - \left. 
    \frac{d\ln \left( k^3 |u_k|^2 \right)}{d \ln k} \right|_{\eta\to 0}
   = 2 \nu - 3 = \frac{2 \epsilon}{1 - \epsilon}.
\end{gather}

We can also calculate that $(1/2) \dot \phi_0^2 / H^2 = \epsilon$ and $\ddot \phi_0/\dot\phi_0 H = -\epsilon$. Then, (\ref{eq:main_eq}) becomes
\begin{align}\label{eq:power_law_K}
  K = \frac{i}{2} \epsilon^2 u_{k_1}^2(\bar{\eta}) 
   \int_{-\infty (1-i\epsilon)}^{\bar{\eta}} d\eta
    \left( a^2 u_{k_1}'^{*2}(\eta)
      + a^2 k_1^2 {u_{k_1}^{*2}}(\eta) \right) + \text{c.c.}\;\;.
\end{align}
Observe that
\begin{align*}
  u'_k(\eta) &= - \frac{c}{\epsilon} 
   \left( \frac{\epsilon}{1 - \epsilon} \right)^{- \frac{1}{1-\epsilon} }
   \exp \left( \frac{i \pi \nu}{2} + \frac{i \pi}{4} \right)
   (-\eta)^{\nu-1} \left[ \nu H_\nu^{(1)}(-k \eta) 
     + (-k \eta) \frac{d\, H_\nu^{(1)}(-k \eta)}{d\,(-k \eta)} \right]\cr
  & = - \frac{c}{\epsilon} 
   \left( \frac{\epsilon}{1 - \epsilon} \right)^{- \frac{1}{1-\epsilon} }
   \exp \left( \frac{i \pi \nu}{2} + \frac{i \pi}{4} \right)
   k (-\eta)^{\nu} H_{\nu-1 }^{(1)}(-k \eta),
\end{align*}
so that
\begin{align*}
  K = -2 c^2 P(k_1) \left\{ i \int^{\infty (1-i\epsilon)}_{\bar{z}} dz \, z
    \left( \left[H^{(2)}_{\nu-1}(z)\right]^2
       + \left[H^{(2)}_{\nu}(z)\right]^2 \right)
     + \text{c.c.} \right\},
\end{align*}
where $z \equiv - i k_1 \eta$.

Using the equations from appendix \ref{sec_ap:properties_of_Hankel_functions}, we find
\begin{gather*}
  \int^{\infty (1-i\epsilon)}_{\bar{z}} dz \, z \left[H^{(2)}_{\nu}(z)\right]^2
    + \text{c.c.} = - \frac{4}{\pi} \nu
  \intertext{so}
  K = - 2 c^2 P(k_1) 
    \left\{ -\frac{4}{\pi} (\nu-1) + - \frac{4}{\pi} \nu \right\}
    = P(k_1) \frac{2 \epsilon}{1 - \epsilon}.
\end{gather*}

Then, as desired,
\begin{align}
  \big\langle  \zeta^3 \big\rangle
   &= (2 \pi)^3 \, \delta^{(3)}\!\big(\,{\textstyle\sum} \mbf_i\big) \, 
    P(k_1) P(k_3)\,
     \frac{2 \epsilon}{1 - \epsilon}\cr
   &= (2 \pi)^3 \, \delta^{(3)}\!\big(\,{\textstyle\sum} \mbf_i\big) \, 
    P(k_1) P(k_3)\, (1-n_S).
\end{align}

\subsection{An exactly scale-invariant model by Starobinsky}
\label{subsec:starob_model}

It's worth noting that, in the previous two models, the slow-roll parameters $\epsilon$ and $\tilde{\eta}$ were constant, which greatly simplified our calculation of $K$ because we could pull out the factors of $\epsilon$ in the first two terms of (\ref{eq:dfn_K}) and because the last term is exactly zero\footnote{Because $\ddot \phi_0/\dot \phi_0 H + (1/ 2) \dot \phi_0^2/ H^2  = -\tilde{\eta}+2\epsilon$.}. Motivated to find an inflation model that predicts a scale-invariant spectrum, Starobinsky in 2005 described another exactly solvable inflation model with varying (not necessarily small) slow-roll parameters \cite{Starobinsky:2005ab}; this model has $1 - n_S = 0$, so that, by the consistency relation, it predicts an exactly vanishing squeezed-limit bispectrum and is therefore an interesting test case. To characterize the model, define $z\equiv a \dot{\phi} / H$ (for this section, we will drop the subscript $0$ on $\phi_0$). His model is essentially specified by assuming $z$ to be of the form
\begin{align*}
  z = -\frac{B}{\eta},
\end{align*}
where $B$ is a positive free-parameter. This is enough to show that\footnote{Note that our definition of $u$ appears different from $u_{\text{St}}$ in \cite{Starobinsky:2005ab} because 1) $u=u_{\text{St}}/z$, and 2) we shifted the phase of $u$ so that $P(k_1) = u_k^2(\text{end of inflation})$.},
\begin{align*}
  u_k=\frac{1}{B} \frac{e^{i k \eta}}{\sqrt{2 k}}
   \left(\frac{1}{k}+i\eta\right).
\end{align*}
For canonical single-field models, the criterion for a mode to freeze outside the horizon is that $k^2\ll z''/z$ because then $\zeta'=1/z^2=H^2/a^2 \dot \phi^2$ \cite{mukhanov1992}. For this model, this implies that $k^2 \ll 2/\eta^2$, so that the natural choice for the end of inflation is when all modes are frozen outside the horizon, i.e. that $\eta=0$.\footnote{In principle, one might object that $\epsilon$ rises above unity slightly before $\eta$ reaches 0, implying that inflation ends earlier. However, it's not clear that $\epsilon$ is a better indicator of whether the horizon size is shrinking.}

If we suppose that inflation ends when $\eta=0$, $P(k) = 1/2 B^2 k^3$ which is exactly scale-invariant, as desired.

For this model,
\begin{align}
  \nonumber V(\phi) &= 3 H(\phi)^2 - 2 H'(\phi),
  \intertext{where}
  H(\phi) &= \frac{1}{B} e^{\phi^2/4}
   \left( \int_{\phi}^\infty e^{-\tilde{\phi}^2/4} d\tilde{\phi} + C\right),
\end{align}
and $C$ is another free parameter. For $C=0$, there is an infinite period of inflation (i.e. $\dot a,\ddot a>0$) that becomes slow-roll as $\phi\to\infty$. For $C$ small and negative, there is a finite period of inflation which gets arbitrarily long as $C\to0$. We will thus focus on these cases.

For this model, it is sometimes useful to use $\phi$ as the dynamical variable instead of $\eta$; for the cases of interest, there is a one-to-one correspondence between these variables. We will indicate derivatives with respect to $\phi$ using hats (e.g. $\hat{H}\equiv dH/d\phi$). We find, for $C$ small and negative,
\newcommand{\etaofphi}
  {\left(\int_{\phi}^\infty e^{-\tilde{\phi}^2/4} d\tilde{\phi} +C\right)}
\begin{align}
  \eta &= - \frac{B}{2 a_0} \etaofphi,\cr
  a&=-a_0 \frac{e^{\phi^2}/4}{B \hat{H}} 
   = 2 a_0 \frac{1}{2 e^{-\phi^2/4} - \phi \etaofphi}
\end{align}
where $a_0 \equiv a(\phi=0)$. 
For reference, note that
\begin{align}
   \frac{1}{2}  \frac{\ddot \phi}{\dot \phi H }
     + \frac{1}{4} \frac{\dot \phi^2}{H^2}
    = -1 -\frac{2}{B} \frac{\hat{H}}{H^2}
    = -1 + \frac{\phi\hat{\eta}}{\eta} + \frac{2 \hat{\eta}^2}{\eta^2}.
\end{align}

For this model, $K$ becomes a fairly involved integral, which is best performed in terms of $\phi$. The integral in $K$ can be evaluated after performing a (somewhat complicated) series of integrations by part:
\begin{align*}
  \int d\eta
    \Bigg[ 
     &\frac{ 1}{ 2 } \frac{ \dot \phi_0^4 }{ H^4} a^2 u_{k_1}'^{*2}
      + \frac{ 1}{ 2 } \frac{ \dot \phi_0^4 }{ H^4} a^2 k_1^2
       {u_{k_1}^{*2}} + 2 \frac{\dot \phi_0^2 }{ H^2 } a^3 
       \frac{d }{ dt} \left( \frac{\ddot \phi_0 }{ \dot \phi_0 H }
        + \frac{ 1 }{ 2} \frac{\dot \phi_0^2 }{ H^2} \right)
       {u_{k_1}'^*} u_{k_1}^*
     \Bigg] = \cr
  &= 
   \int d\phi\; \frac{e^{2 i k_1 \eta } }{4 k_1 \eta ^4} 
     \Big[\eta ^2 \hat{\eta } \left(4-\phi ^2\right)-8 \phi  \eta  \hat{\eta }^2-12 \hat{\eta }^3+4 i k_1 \eta
 \left(-\eta ^2 \hat{\eta }+\eta  \phi  \hat{\eta }^2+2 \hat{\eta }^3\right)\Big]=\cr
    &=-\frac{e^{2 i k_1 \eta }}{2 k_1 \eta } \left(1-\frac{\phi  \hat{\eta }}{\eta }-\frac{2 \hat{\eta }^2}{\eta ^2}\right).
\end{align*}
There is a slight subtlety when evaluating this quantity at $\eta \to -\infty (1-i \epsilon)$, for the lower bound of integration in $K$, because in the Starobinsky model, $\eta(\phi)$ doesn't limit to $-\infty$, and in fact, the model behaves unusually for $\phi$ somewhat less than zero. However, one can imagine that for $\phi \approx 0$, the model merges with another inflation model or a static universe. Then, we just let $\eta$ go to $-\infty$ and the $e^{2 i k \eta}$ factor will make the integrand go to zero.

Thus,
\begin{align*}
  K &\equiv i u_{k_1}^2 (\bar{\eta}) \int_{-\infty (1-i \epsilon)}^{\bar{\eta}}d\eta\ldots + \text{c.c.} = 
   \frac{1}{B^2 k_1^3} 
   \left.\left(1-\frac{\phi  \hat{\eta }}{\eta }-\frac{2 \hat{\eta }^2}{\eta ^2}\right)
     \right|_{\eta\to\bar{\eta}},
\end{align*}
and the term in brackets in (\ref{eq:main_eq}) is
\begin{align*}
  \left. \left| u_{k_1}\right|^2\! \;
    2 \left( \frac{\ddot \phi }{ \dot \phi H } 
      + \frac{ 1 }{ 2} \frac{\dot \phi^2 }{ H^2} \right) + K \;\right|_{\eta\to\bar{\eta}}
     &= \left. \frac{1}{B^2 k_1} 
       \left(-\eta ^2+\eta  \phi  \hat{\eta }+2 \hat{\eta }^2\right) \right|_{\eta\to\bar{\eta}}\cr
      &\stackrel{\bar{\eta}\to0}{=} \frac{2}{B^2 k_1} \hat{\eta}^2
       = P(k_1) \frac{B^2 k_1^2}{a_0^2} e^{-\phi_{\text{end}}^2/2},
\end{align*}
where  $\phi_{\text{end}}\equiv\phi(\eta=0)$ (i.e. $\phi_{\text{end}} = 2 \operatorname{erfc}^{-1}\left(-\frac{c}{\sqrt{\pi}}\right)$, $\operatorname{erfc}$ is the complementary error function), yielding
\begin{align}\label{eq:starob-f_nl}
  f_{\text{NL}}= (B^2 k_1^2/a_0^2) e^{-\phi_{\text{end}}^2/2}.
\end{align}

Since one expects to find $f_{\text{NL}} = 1-n_S = 0$ for the Starobinsky model, this result initially appears confusing. However, as we will explain, $f_{\text{NL}}$ is non-zero only to the extent that some of the conditions of the consistency relation are violated. First, we will discuss the factor $B^2 k_1^2/a_0^2$. As mentioned earlier, the comoving horizon size is approximately $1/(1/\eta)=\eta$ so that, at $\phi=0$, the horizon size is $|\eta(\phi=0)|\approx B/2 a_0$, since $C$ is taken to be small. Thus, the ratio $B k_1 / a_0$ specifies how close a mode $k_1$ is to the horizon size at the start of inflation. Since (for $C\neq0$) the Starobinsky model has a finite inflationary period, this ratio also determines how long the mode will spend outside the horizon; note that, formally, all modes cross the horizon as $\eta\to 0$ but some of them will spend only extremely short times outside the horizon. The second factor $e^{-\phi_{\text{end}}^2/2}$ has a similarly straightforward interpretation: as $C\to 0$,  $\phi_\text{end}\to \infty$ and the inflationary period gets arbitrarily long, so all modes spend an increased time outside the horizon. Both of these factors are reasonable since the consistency relation should only hold if the perturbations are outside the horizon ``long enough'' to freeze out and to stop varying; our result is interesting in that it shows quantitatively how $f_{\text{NL}}$ approaches the consistency relation prediction as the parameters approach the conditions it assumes.

\section{Discussion}
\label{sec:discussion}

In this paper, we developed a technique for calculating squeezed-limit correlation functions. We applied it to the single-field bispectrum with a canonical kinetic term and arrived at (\ref{eq:main_eq}), a formula for the squeezed-limit bispectrum which relies on no slow-roll approximation. We first verified that our technique matched the known squeezed-limit result for slow-roll inflation. We then performed the calculation for power-law inflation to arbitrary order in slow-roll and explicitly verified that it matched the consistency relation prediction. Finally, in the Starobinsky model, we again saw that the consistency relation held, though we found corrections that depended on the ratio of the size of the $k_1$ mode and the horizon size and on the length of the inflationary period. Thus, one has to be careful that the assumptions of the consistency relation are satisfied before expecting it to hold. Even for a mode that leaves the horizon, if the subsequent inflationary period is sufficiently short, it could have an enhanced squeezed-limit bispectrum. 

Our technique can also be applied to cases with non-standard kinetic terms. This was done in \cite{RenauxPetel:2010ty}; following our method, the author derived the formula for the squeezed-limit bispectrum for any single-field model where the action for the scalar is given by $P(X,\phi)$, where $X\equiv-(1/2)g^{\mu\nu}\partial_\mu\phi\partial_\nu\phi$. He verified the validity of the consistency relation for power-law inflation where the speed of sound $c_s\neq 1$, as well as for second-order slow-roll inflation.

Besides providing explicit verification for the consistency relation, our technique provides a method of calculating $\big\langle \zeta_{\mbf[k]_1} \zeta_{\mbf[k]_2} \big\rangle_{\zeta_{\mbf[k]_3}}$, an important part of the consistency relation's derivation. Furthermore, as compared to the proof of the consistency relation, it is done purely in Fourier space, i.e. it provides a complementary approach.

We would also like to point out that our formalism could be useful for calculating other squeeze-limit correlation functions, since our primary assumption is that wavemodes which stretch far enough outside the horizon turn classical. In particular, our technique could be adapted for calculating the squeezed limit trispectrum or multi-field inflation bispectrum.

\acknowledgments

The authors particularly appreciate many enlightening conversations with Joel Meyers, as well as discussions about several aspects of the paper with Donghui Jeong. Yuki Watanabe was kind enough to point out several notable corrections. We are also grateful for helpful ideas from Sonia Paban and Steven Weinberg. We want to thank the organizers of the Yukawa International Seminar 2010 (YKIS2010) and Gravitation and Cosmology 2010 (YITP-T-10-01), where this work was completed; we appreciate both their warm hospitality and the stimulating presentations. This work is supported in part by NSF grant PHY-0758153.

\appendix
\allowdisplaybreaks[1]
\section[Some explicit calculations]{Some explicit calculations}
\label{sec_ap:in_depth_calcs}

\subsection[Field redefinition calculation from subection \simpref{subsec:fld_rdfnn}]{Field redefinition calculation from subsection \protect\ref{subsec:fld_rdfnn}}
\label{subsec_ap:fld_rdfnn_calc}

If $\zeta_S(t, \mbf[x]) = \zeta_N(t,\mbf[x]) + \beta(t) \zeta_L(t,\mbf[x]) \zeta_N(t,\mbf[x])$, then
\begin{align*}\label{crrn-fcn}
  \big\langle \zeta_{S,\mbf_1} & (t)  \zeta_{S,\mbf_2}(t)  \big\rangle = \cr
    &= \int d^3 x_1 d^3 x_2 
      \left\langle \zeta_{S}(\mbf[x]_1) \zeta_{S}(\mbf[x]_2) \right\rangle
      e^{-i(\mbf_1 \cdot \mbf[x]_1+\mbf_2 \cdot \mbf[x]_2)} \approx \cr
   &\approx \left\langle \zeta_{N,\mbf_1} \zeta_{N,\mbf_2} \right\rangle 
 \int d^3 x_1 d^3 x_2 
    \left\langle \zeta_{N}(\mbf[x]_1) \zeta_{N}(\mbf[x]_2) \right\rangle \Big[
      \beta \, \zeta_L(\mbf[x]_1)
          + \beta \, \zeta_L(\mbf[x]_2) \Big]
      e^{-i(\mbf_1 \cdot \mbf[x]_1+\mbf_2 \cdot \mbf[x]_2)} \approx\cr
  & \approx \left\langle \zeta_{N,\mbf_1} \zeta_{N,\mbf_2} \right\rangle + \cr
   &~~~~+ 2 \, \beta \, 
     \int \prod_i \frac{d^3 q_i}{(2 \pi)^3} d^3 x_1 d^3 x_2 \, 
    \zeta_{L,\mbf[q]_3 }
     \underbrace{\left\langle \zeta_{N,\mbf[q],1} \zeta_{N,\mbf[q],2} \right\rangle}
      _{=(2\pi)^3 \ft (q_1) \, \delta(\mbf[q]_1 + \mbf[q]_2)} 
    e^{i \left( (\mbf[q]_3 + \mbf[q]_1 - \mbf_1)\cdot\mbf[x]_1
      + (\mbf[q]_2 - \mbf_2)\cdot\mbf[x]_2 \right)} = \cr
  &= \left\langle \zeta_{N,\mbf_1} \zeta_{N,\mbf_2} \right\rangle 
    +  2 \, \beta(t) \,
      \zeta_{L,\mbf_1 + \mbf_2}(t)
      \, \ft\left[ \big\langle \zeta^2(\Delta x,t) \big\rangle \right]\!(k_1)\,
\end{align*}
matching the result from earlier. As noted earlier in a footnote,
$\ft[f(\mbf[x])](\mbf)$ is defined as
\begin{align*}
  \ft[f(\mbf[x])](\mbf)
    \equiv \int d^3x f(\mbf[x]) e^{-i \mbf\cdot\mbf[x]}.
\end{align*}

\subsection[In-in calculation from subsection \simpref{subsec:in_in_formalism}]{In-in calculation from subsection \protect\ref{subsec:in_in_formalism}}
\label{subsec_ap:in_in_calc}

First, let us calculate from (\ref{eq:sint}), using (\ref{eq:expd_zet}) and (\ref{eq:qntz_zet_N}),
\begin{align}
 &\begin{aligned}
  H_I(t) = - L_\text{int}
 \end{aligned}\cr
 &\begin{aligned}
  \phantom{H_I(t)}
   =- \int & \frac{d^3 q_1 d^3 q_2}{(2\pi)^6} \zeta_{L,-\mbf[q]_1 -\mbf[q]_2}\times\cr
    &\times\left[ \left(\frac{1}{4} \frac{ \dot \phi_0^4 }{ H^4} 
    -\frac{ 1 }{ 16} \frac{ \dot \phi_0^6 }{ H^6} \right) 
   a^3 \dot \zeta_{\mbf[q]_1} \dot \zeta_{\mbf[q]_2}
   - \frac{ 1}{ 4 } \frac{ \dot \phi_0^4 }{ H^4} a 
    \mbf[q]_1 \cdot \mbf[q]_2 \zeta_{\mbf[q]_1} \zeta_{\mbf[q]_2} \right.+ \cr
  & \phantom{\times\Big[\Big(}
   \left.  + \frac{ 1 }{ 16} \frac{ \dot \phi_0^6 }{ H^6 } 
     a^3 \frac{(\mbf[q]_1 \cdot \mbf[q]_2)^2}{q_1^2 q_2^2}
     \dot \zeta_{\mbf[q]_1} \dot \zeta_{\mbf[q]_2}
    + \frac{\dot \phi_0^2 }{ H^2 } a^3 
   \frac{d }{ dt} \left( \frac{\ddot \phi_0 }{ \dot \phi_0 H } 
     + \frac{ 1 }{ 2} \frac{\dot \phi_0^2 }{ H^2} \right) \dot \zeta_{\mbf[q]_1} \zeta_{\mbf[q]_2}
    \right],
 \end{aligned}\cr
\end{align}
where we have neglected the $\dot \zeta_N \partial_i \zeta_N \partial_i \partial^{-2} \dot \zeta_L$ term because it will not contribute in the squeezed limit (this is further discussed at the end of this subsection). 

Also,
\begin{align}
  \big\langle\, 0 \big| \big[&  \zeta_{\mbf[k]_1}(\bar{t}) \zeta_{\mbf[k]_2}(\bar{t}), 
    \zeta_{\mbf[k]_1}(t) \zeta_{\mbf[k]_2}(t) \big] \big| 0 \, \big\rangle = \cr
  &\begin{aligned}
  = \Big\langle\, 0 \Big| \Big(u_{k_1} &(\bar{t}) \, a_{\mbf[k]_1} 
    + u^*_{k_1}(\bar{t}) \, a^\dagger_{-\mbf[k]_1}\Big)
  \Big(u_{k_2}(\bar{t}) \, a_{\mbf[k]_2} 
    + u^*_{k_2}(\bar{t}) \, a^\dagger_{-\mbf[k]_2}\Big) \times\cr
  &\times\Big(u_{q_1}(t) \, a_{\mbf[q]_1} 
     + u^*_{q_1}(t) \, a^\dagger_{-\mbf[q]_1}\Big)
    \Big(u_{q_2}(t) \, a_{\mbf[q]_2} 
     + u^*_{q_2}(t) \, a^\dagger_{-\mbf[q]_2}\Big)
   \Big| 0 \, \Big\rangle-
  \end{aligned}\cr
  &\begin{aligned}
  \phantom{=}- \Big\langle\, 0 \Big| 
   \Big(u_{q_1}&(t) \, a_{\mbf[q]_1} 
     + u^*_{q_1}(t) \, a^\dagger_{-\mbf[q]_1}\Big)
   \Big(u_{q_2}(t) \, a_{\mbf[q]_2} 
     + u^*_{q_2}(t) \, a^\dagger_{-\mbf[q]_2}\Big)\times\cr
  &\times\Big(u_{k_1} (\bar{t}) \, a_{\mbf[k]_1} 
    + u^*_{k_1}(\bar{t}) \, a^\dagger_{-\mbf[k]_1}\Big)
   \Big(u_{k_2}(\bar{t}) \, a_{\mbf[k]_2} 
    + u^*_{k_2}(\bar{t}) \, a^\dagger_{-\mbf[k]_2}\Big)  \Big| 0 \, \Big\rangle=
  \end{aligned}\cr
  &\begin{aligned}
    =(2 & \pi)^6\Big(
     u(k_1,t) u(k_2,t)
      (u(q_1,t') u(q_2,t'))^* - \text{c.c.}\Big) \times\cr
     &\times\left[
      \delta(\mbf[k]_1 +\mbf[q]_1) \delta(\mbf[k]_2 +\mbf[q]_2)
      + \delta(\mbf[k]_1 +\mbf[q]_2) \delta(\mbf[k]_2 +\mbf[q]_1)
    \right]
  \end{aligned}\cr
\end{align}
There are also variants where the $\zeta$ terms with $q$ indices have time derivatives, for example where $\zeta_{\mbf[q]_1}(t)$ is replaced by $\dot \zeta_{\mbf[q]_1}(t)$; in these cases, we simply apply the time derivative to the corresponding mode function, so that $u_{\mbf[q]_1}(t) \to \dot u_{\mbf[q]_1}(t)$.

\newcommand{\textone}[0]{asfd}

Then, 
\begin{align}
  \big\langle \zeta_{N,\mbf[k]_1} &\zeta_{N,\mbf[k]_2} \big\rangle_{\zeta_{\mbf[k]_3} }=\cr
  &=i \int_{t_0}^t dt' \frac{d^3 q_1 d^3 q_2}{(2\pi)^6} 
      \zeta_{L,-\mbf[q]_1 -\mbf[q]_2}\times\cr
    &\phantom{\textone}\times\Bigg\{
     \left(\frac{1}{4} \frac{\dot\phi_0^4}{H^4} 
      -\frac{1}{16} \frac{\dot \phi_0^6 }{ H^6} \right)(t') a(t')^3
     \left\langle0 \left|
        \left[\zeta_{\mbf[k]_1}\zeta_{\mbf[k]_2}, 
          \dot \zeta_{\mbf[q]_1}\dot \zeta_{\mbf[q]_2}\right]
         \right|0\right\rangle - \cr
    &\phantom{\textone}~~~~~ - \frac{1}{4} \frac{\dot \phi_0^4}{ H^4}(t') a(t') \mbf[q]_1 \cdot \mbf[q]_2
      \left\langle0 \left|         
       \left[\zeta_{\mbf[k]_1}\zeta_{\mbf[k]_2}
         , \zeta_{\mbf[q]_1}\zeta_{\mbf[q]_2} \right]
        \right|0\right\rangle + \cr
    &\phantom{\textone}~~~~~ + \frac{1}{ 16 } \frac{\dot \phi_0^6 }{ H^6}(t') a^3(t') 
     \frac{(\mbf[q]_1 \cdot \mbf[q]_2)^2}{q_1^2 q_2^2}
      \left\langle0 \left|         
       \left[\zeta_{\mbf[k]_1}\zeta_{\mbf[k]_2}
         , \dot \zeta_{\mbf[q]_1}\dot \zeta_{\mbf[q]_2} \right]
        \right|0\right\rangle + \cr
    &\phantom{\textone}~~~~~ + \frac{\dot \phi_0^2 }{ H^2 }(t') a(t')^3 
      \frac{d }{ dt} \left[ \frac{\ddot \phi_0 }{ \dot \phi_0 H }
     + \frac{1 }{ 2} \frac{\dot \phi_0^2 }{ H^2} \right] (t')
      \left\langle0 \left| \left[\zeta_{\mbf[k]_1}\zeta_{\mbf[k]_2}, 
            \dot \zeta_{\mbf[q]_1}\zeta_{\mbf[q]_2} \right]
         \right|0\right\rangle \Bigg\} \approx \cr
 &\begin{aligned}
   \approx i \zeta_{L,\mbf[k]_1 + \mbf[k]_2}
    \Bigg\{\int_{t_0}^t dt' &2 \left(\frac{1}{4} \frac{\dot \phi_0^4}{H^4} 
      -\frac{1}{16} \frac{\dot \phi_0^6 }{ H^6} \right)\!\!(t') a(t')^3
     \; u(k_1,t) u(k_2,t) \big(\dot u(k_1,t') \dot u(k_2,t') \big)^*-\cr
     & -2\,\frac{1}{4} \frac{\dot \phi_0^4}{ H^4}(t')a(t') \mbf[k]_1 \cdot \mbf[k]_2
      \; u(k_1,t) u(k_2,t) \big(u(k_1,t') u(k_2,t') \big)^* \cr
    & + 2 \, \frac{1}{ 16 } \frac{\dot \phi_0^6 }{ H^6}(t') a^3(t') 
     \frac{(\mbf[k]_1 \cdot \mbf[k]_2)^2}{k_1^2 k_2^2}        
      \; u(k_1,t) u(k_2,t) \big(\dot u(k_1,t') \dot u(k_2,t') \big)^*\cr
     &\begin{aligned} +& \frac{\dot \phi_0^2 }{ H^2 }(t') a(t')^3 
      \frac{d }{ dt} \left[ \frac{\ddot \phi_0 }{ \dot \phi_0 H }
        + \frac{1 }{ 2} \frac{\dot \phi_0^2 }{ H^2} \right] (t') \times\cr
      & \times 2 u(k_1,t) u(k_2,t)
          \big(\dot u(k_1,t') u(k_2,t')\big)^* 
       \quad - \;\;\text{c.c.}\Bigg\}\end{aligned}
  \end{aligned}\cr
 &\overset{\mbf_1 \approx \mbf_2}{=} \zeta_{L,\mbf[k]_1 + \mbf[k]_2} K \; ,
\end{align}
where $K$ is defined in (\ref{eq:main_eq}).

Now let us return to the piece we neglected earlier:
\begin{align}
   L_\text{int}(t)  \subset \int - \frac{ \dot \phi_0^4 }{ 2 H^4} a^3 
    \dot \zeta_S \partial_i \zeta_S \partial_i \partial^{-2} \dot \zeta_L.
\end{align}

This contributes\begin{align}
  H_I(t) \subset  
   \int \frac{d^3 q_1 d^3 q_2}{(2\pi)^6} \frac{\dot \phi_0^4 }{ 2 H^4 } a^3 
      \dot \zeta_{L,- \mbf[q]_1 - \mbf[q]_2} 
      \frac{\mbf[q]_2 \cdot (- \mbf[q]_1 - \mbf[q]_2)}{(-\mbf[q]_1 - \mbf[q]_2)^2} 
      \dot \zeta_{\mbf[q]_1} \zeta_{\mbf[q]_2}\, ,
\end{align}
so that 
\begin{align}
  \big\langle \zeta_{\mbf[k]_1} & \zeta_{\mbf[k]_2} \big\rangle \subset \cr
 &\subset - i \int_{t_0}^t dt' \frac{d^3 q_1 d^3 q_2}{(2\pi)^6} 
      \dot \zeta_{L,- \mbf[q]_1 - \mbf[q]_2}
      \frac{\dot \phi_0^4 }{ 2 H^4 } a^3 
      \frac{\mbf[q]_2 \cdot (- \mbf[q]_1 - \mbf[q]_2)}{(-\mbf[q]_1 - \mbf[q]_2)^2}
       \left\langle0 \left| \left[\zeta_{\mbf[k]_1}\zeta_{\mbf[k]_2}, 
            \dot \zeta_{\mbf[q]_1}\zeta_{\mbf[q]_2} \right]
         \right|0\right\rangle = \cr
 &\begin{aligned}
  =- i \int_{t_0}^t dt'
     & \frac{\dot \phi_0^4 }{ 2 H^4 }(t') a(t')^3 
      \dot \zeta_{L,\mbf[k]_1 + \mbf[k]_2}
      \frac{- \mbf[k]_1 \cdot (\mbf[k]_1 + \mbf[k]_2)}{(\mbf[k]_1 + \mbf[k]_2)^2} \times \cr
      & \times \Big[ 
               u(k_1,t) u(k_2,t) \big(\dot u(k_1,t') u(k_2,t')\big)^* 
      - \text{c.c.}\Big] + \mbf[k]_1 \leftrightarrow \mbf[k]_2 = 
 \end{aligned}\cr
 &\begin{aligned}
  = i \int_{t_0}^t dt'
     & \frac{\dot \phi_0^4 }{ 2 H^4 }(t') a(t')^3 
      \dot \zeta_{L,\mbf[k]_1 + \mbf[k]_2}
      \Big[ 
         u^2(k_1,t) \left(\dot u(k_1,t') u(k_1,t') \right)^*
      - \text{c.c.}\Big]\,.
 \end{aligned}
\end{align}

However, this is the same contribution we would get from a term
\begin{align}
  L_\text{int}(t) \subset 
    \int - \frac{ \dot \phi_0^4 }{ 4 H^4} a^3 \dot \zeta_S \zeta_S \dot \zeta_L,
\end{align}
which is clearly non-dominant in the squeezed limit.

\section[Properties of Hankel functions used in subsection \simpref{subsec:power_law_iftn}]{Properties of Hankel functions used in subsection \protect\ref{subsec:power_law_iftn}}
\label{sec_ap:properties_of_Hankel_functions}
Here, we collect the properties of Hankel functions that are used in our calculations (these can be found in any standard reference, e.g. \cite{grad&ryz}). [Equations that do not specify $H^{(1)}$ or $H^{(2)}$ apply to both and arguments shown as $z$ can be complex while arguments shown as $x$ are real.]
\begin{gather}
   H_\nu^{(1)}(x) = H_\nu^{(2)*}(x),\\
   z \frac{d}{dz} H_\nu(z) - \nu H_\nu(z) = zH_{\nu + 1}(z),\\
   \int dx \,x \left[ H_\nu(x) \right]^2 = \frac{x^2}{2}
     \left[ \left[H_\nu(z)\right]^2 - H_{\nu-1}(x) H_{\nu+1}(x)\right]\\
   H^{\binom{1}{2}}_{\nu}(z)\to \frac{1}{\Gamma(\nu+1)} 
    \genfrac{(}{)}{}{}{x}{2}^\nu
     \pm i \frac{\Gamma(\nu)}{\pi} \left( \frac{2}{x} \right)^\nu
    \qquad \text{small real }x\\
   H^{\binom{1}{2}}(z) \to \sqrt{\frac{2}{\pi z}}
      e^{\pm i \left(z - \frac{\pi}{2} \nu - \frac{\pi}{4} \right)}
    \qquad \text{large $|z|$, $|\arg z|<\pi$}
\end{gather}

\bibliographystyle{JHEP}	
\bibliography{paper-bib}

\end{document}